\documentclass[envcountsect,envcountsame]{llncs}

\usepackage{latexsym}
\usepackage{amssymb}
\usepackage{amsmath}
\usepackage{float}
\usepackage{framed}
\usepackage{qip}
\usepackage{bbm}  
\usepackage{tabularx}
\usepackage{xspace}

\allowdisplaybreaks
\newcounter{itm}

\newcommand{\from}{\leftarrow}

\newcommand{\set}[1]{\{#1\}}

\newcommand{\bias}[1][\alpha]{\mathrm{bias}_{#1}}

\newcommand{\Norm}[1]{|\!|\!|#1|\!|\!|_2}

\newcommand*{\dens}[1]{\mathcal{P}(#1)}  

\newcommand{\eps}{\varepsilon}

\newcommand{\E}{\mathbbm{E}}   

\newcommand{\K}{{\cal K}}

\renewcommand*{\H}{\operatorname{H}}   

\newcommand{\nord}{{\circ}}

\newcommand{\Hoo}[1][]{\mathrm{H}_{\infty}^{#1}}

\newcommand{\F}{\frak{F}}

\newcommand*{\cH}{\mathcal{H}}
\newcommand*{\cI}{\mathcal{I}}

\newcommand*{\cX}{\mathcal{X}}

\newcommand*{\assign}{\ensuremath{\kern.5ex\raisebox{.1ex}{\mbox{\rm:}}\kern -.3em =}}

\renewcommand*{\id}{\mathbbm{1}}

\newcommand{\etal}{{\it et al.}}

\newcommand{\delete}[1]{}
\newcommand{\remove}[1]{}

\def\version$#1,v #2 #3/#4/#5 #6${#2 (#5-#4-#3)}

\title{
Randomness Extraction via $\delta$-Biased Masking \\ in the Presence of a Quantum~Attacker
}

\author{
Serge Fehr\thanks{Supported by a Veni grant from the Dutch Organization for Scientific Research (NWO).} 
\and 
Christian Schaffner\thanks{Supported by the EU projects SECOQC and QAP IST 015848
  and a NWO Vici grant 2004-2009.}
}

\tocauthor{
 Serge Fehr (CWI Amsterdam, The Netherlands),
 Christian Schaffner (CWI Amsterdam, The Netherlands)
 }

\institute{
CWI\thanks{Centrum voor Wiskunde en Informatica, the national research institute for mathematics and computer science in the Netherlands. } Amsterdam, The Netherlands \\
\email{\{S.Fehr,C.Schaffner\}@cwi.nl}
}

\begin{document}

\maketitle

\begin{abstract}
  Randomness extraction is of fundamental importance for
  information-theoretic cryptography. It allows to transform a raw key
  about which an attacker has some limited knowledge into a
  fully secure random key, on which the attacker has essentially no
  information.
  Up to date, only very few randomness-extraction techniques are known
  to work against an attacker holding quantum information on the raw key.  
  This is very much in contrast to the
  classical (non-quantum) setting, which is much better understood and
  for which a vast amount of different techniques are known and proven to work. 
  
  We prove a new randomness-extraction technique, which is known to work in the 
  classical setting, to be secure against a quantum attacker as well.
  Randomness extraction is done by xor'ing a so-called $\delta$-biased
  mask to the raw key. 
  Our result allows to extend the classical applications of this extractor to the quantum setting. 
  We discuss the following two applications. 
  We show how to encrypt a long message with a short key,
  information-theoretically secure against a quantum attacker,
  provided that the attacker has enough quantum uncertainty on the
  message. This generalizes the concept of entropically-secure
  encryption to the case of a quantum attacker.
  As second application, we show how to do error-correction
  without leaking partial information to a quantum attacker. Such a technique is
  useful in settings where the raw key may contain errors, since standard
  error-correction techniques may provide the attacker with
  information on, say, a secret key that was used to obtain the raw
  key.
\end{abstract}

\section{Introduction}

Randomness extraction allows to transform a raw key $X$ about which an
attacker has some limited knowledge into a fully secure random key
$S$. It is required that the attacker has essentially no
information on the resulting random key $S$, no matter what kind of
information he has about the raw key $X$, as long as his uncertainty
on $X$ is lower bounded in terms of a suitable entropy measure.
One distinguishes between extractors which use a private seed (preferably as small as possible)~\cite{TaS96}, and, what is nowadays called {\em strong} extractors, which only use public coins~\cite{ILL89,NZ93}. In the context of cryptography, the latter kind of randomness extraction is also known as privacy amplification~\cite{BBCM95}. 
Randomness-extraction techniques play an important role in various areas of theoretical computer science. In cryptography, they are at the core of many constructions
in information-theoretic cryptography, but they also proved to be useful in the computational setting. 
As such, there is a huge amount of literature on randomness extraction, and there exist various techniques which are
optimized with respect to different needs; we refer to Shaltiel's survey~\cite{Shaltiel02} for an informative overview on classical and recent results. 

Most of these techniques, however, are only guaranteed to work in a non-quantum
setting, where information is formalized by means of classical information theory. 
In a quantum setting, where the attacker's
information is given by a quantum state, our current understanding
is much more deflating.  Renner and K\"onig~\cite{RK05} have shown that privacy
amplification via universal$_2$ hashing is secure against quantum adversaries.  
And, K\"onig and Terhal~\cite{KT06} showed
security against quantum attackers for certain extractors, namely for one-bit-output strong extractors, as well as for strong extractors which work by extracting bit wise via one-bit-output strong extractors. 
Concurrent to our work, Smith has shown recently that Renner and K\"onig's result generalizes to {\em almost}-universal hashing, i.e., that  Srinivasan-Zuckerman extractors remain secure against quantum adversaries~\cite{Smith07privcom}. 
On the negative side, Gavinsky \etal\ recently showed that there exist (strong) extractors that are secure against classical attackers, but which become completely insecure against quantum attackers~\cite{GKKRW07}. Hence, it is not only a matter of lack of proof, but in fact classical extractors may turn insecure when considering {\em quantum} attackers. 

We prove a new randomness-extraction technique to be secure
against a quantum attacker. It is based on the concept of \emph{small-biased spaces}, see e.g. \cite{NN90}. Concretely, randomness extraction is done by xor'ing
the raw key $X \in \set{0,1}^n$ with a $\delta$-biased mask $A \in \set{0,1}^n$, chosen privately
according to some specific distribution, where the distribution may be
chosen publicly from some family of distributions. Roughly, $A$
(or actually the family of distributions) is $\delta$-biased,
if any non-trivial parity of $A$ can only be guessed with
advantage~$\delta$. We prove that if $A$ is $\delta$-biased, then the bit-wise xor $X \oplus A$ is $\eps$-close to random and independent of the attacker's
quantum state with $\eps = \delta\cdot2^{(n-t)/2}$, where $t$ is the attacker's quantum collision-entropy in
$X$. Thus, writing $\delta = 2^{-\kappa}$, the extracted key $X \oplus
A$ is essentially random as long as $2\kappa$ is significantly larger
than $n-t$. 
Note that in its generic form, this randomness extractor uses public coins, namely the choice of the distribution, {\em and} a private seed, the sampling of $A$ according to the chosen distribution. Specific instantiations though, may lead to standard extractors with no public coins (as in Section~\ref{sec:AppI}), or to a strong extractor with no private seed (as in Section~\ref{sec:AppII}). The proof of the new
randomness-extraction result combines quantum-information-theoretic techniques developed by
Renner~\cite{Renner05,RK05} and techniques from Fourier
analysis, similar to though slightly more involved than those used in~\cite{AS04}. 

We would like to point out that the particular extractor we consider,
$\delta$-biased masking, is well known to be secure against {\em
non}-quantum attackers. Indeed, classical security was shown by
Dodis and Smith, who also suggested useful
applications~\cite{DS05,DS05tcc}. Thus, our main contribution is the
{\em security analysis} in the presence of a {\em quantum} attacker. Our
positive result not only contributes to the general problem of
the security of extractors against quantum attacks, but it is particularly
useful in combination with the classical applications of
$\delta$-biased masking where it leads to interesting new results in the
quantum setting. We discuss these applications and the arising new
results below.

The first application is entropically secure encryption~\cite{RW02,DS05tcc}.
An encryption scheme is entropically secure if the ciphertext gives
essentially no information away on the plaintext (in an
information-theoretic sense), provided that the attacker's a priori
information on the plaintext is limited. Entropic security allows to
overcome Shannon's pessimistic result on the size of the key for
information-theoretically secure encryption, in that a key of size
essentially $\ell \approx n - t$ suffices to encrypt a plaintext of
size $n$ which has $t$ bits of entropy given the attacker's a priori
information. This key size was known to suffice for a non-quantum
adversary~\cite{RW02,DS05tcc}. By our analysis, this result
carries over to the setting where we allow the attacker
to store information as quantum states: a key of size essentially
$\ell \approx n - t$ suffices to encrypt a plaintext of size $n$
which has $t$ bits of (min- or collision-) entropy given the attacker's
quantum information about the plaintext.

Note that entropic security in a quantum setting was also considered
explicitly in~\cite{Des07} and implicitly for the task of approximate
quantum encryption~\cite{AS04,KN05,DN06}. However, all these results
are on encrypting a {\em quantum} message into a quantum ciphertext on
which the attacker has limited {\em classical} information (or none at
all), whereas we consider encrypting a {\em classical} message into a
classical ciphertext on which the attacker has limited {\em quantum}
information. Thus, our result in quantum entropic security is in that
sense orthogonal. 
As a matter of fact, the results in~\cite{AS04,KN05,DN06,Des07} about
randomizing quantum states can also be appreciated as extracting 
``quantum randomness" from a quantum state on which the attacker has
limited {\em classical} information. Again, this is
orthogonal to our randomness-extraction result which allows to extract
classical randomness from a {\em classical} string on which the
attacker has limited {\em quantum} information. 
In independent recent work, Desrosiers and Dupuis showed that one can combine
techniques to get the best out of both: they showed that $\delta$-biased masking (as used in \cite{AS04}) allows to extract
``quantum randomness" from a {\em quantum} state on which the attacker has
limited {\em quantum} information. This in particular implies our result. 

The second application is in the context of private error-correction.
Consider a situation where the raw key $X$ is obtained by Alice and
Bob with the help of some (short) common secret key $K$, and where the
attacker Eve, who does not know $K$, has high entropy on $X$. Assume
that, due to noise, Bob's version of the raw key $X'$ is slightly
different from Alice's version $X$. Such a situation may for instance
occur in the bounded-storage model or in a quantum-key-distribution
setting. Since Alice and Bob have different versions of the raw key,
they first need to correct the errors before they can extract (by
means of randomness extraction) a secure key $S$ from $X$.  However,
since $X$ and $X'$ depend on $K$, standard techniques for correcting
the errors between $X$ and $X'$ leak information not only on $X$ but
also on $K$ to Eve, which prohibits that Alice and Bob can re-use $K$
in a future session. In the case of a non-quantum attacker, Dodis and
Smith showed how to do error-correction in such a setting without
leaking information on $K$ to Eve~\cite{DS05}, and thus that $K$ can
be safely re-used an unlimited number of times. We show how our
randomness-extraction result gives rise to a similar way of doing
error correction without leaking information on $K$, even if Eve holds
her partial information on $X$ in a quantum state.  Such a
private-error-correction technique is a useful tool in various
information-theoretic settings with a quantum adversary. Very
specifically, this technique has already been used as essential
ingredient to derive new results in the bounded-(quantum)-storage
model and in quantum key distribution~\cite{DFSS07}.

The paper is organized as follows. We start with some
quantum-information-theoretic notation and
definitions. The new
randomness-extraction result is presented in Section~\ref{sec:main}
and proven in Section~\ref{sec:proof}. The two applications discussed
are given in Sections~\ref{sec:AppI} and~\ref{sec:AppII}.

\section{Preliminaries}
\subsection{Notation and Terminology}

A {\em quantum system} is described by a complex Hilbert space $\cH_A$
(in this paper always of finite dimension). The {\em state} of the
system is given by a {\em density matrix}: a positive semi-definite
operator $\rho_A$ on $\cH_A$ with trace $\tr(\rho_A) = 1$. We write
$\dens{\cH_A}$ for the set of all positive semi-definite operators on
$\cH_A$, and we call $\rho_A \in \dens{\cH_A}$ {\em normalized} if it
has trace $1$, i.e., if it is a density matrix.  For a density matrix
$\rho_{AB} \in \dens{\cH_A \otimes \cH_B}$ of a composite quantum
system $\cH_A \otimes \cH_B$, we write $\rho_{B} = \tr_A(\rho_{AB})$
for the state obtained by tracing out system $\cH_{A}$.  A density
matrix $\rho_{XB} \in \dens{\cH_X \otimes \cH_B}$ is called {\em
  classical on $\cH_X$ with $X \in {\cal X}$}, if it is of the form
$\rho_{XB} = \sum_x P_X(x) \proj{x}\otimes \rho_B^x$ with normalized
$\rho_B^x \in \dens{\cH_B}$, where $\set{\ket{x}}_{x \in \cal X}$
forms an orthonormal basis of $\cH_X$. Such a density matrix
$\rho_{XB}$ which is classical on $\cH_X$ can be viewed as a random
variable $X$ with distribution $P_X$ together with a family
$\set{\rho_B^x}_{x \in \cal X}$ of {\em conditional density matrices},
such that the state of $\cH_B$ is given by $\rho_B^x$ if and only if
$X$ takes on the value $x$. We can introduce a new random variable $Y$
which is obtained by ``processing" $X$, i.e., by extending the
distribution $P_X$ to a consistent joint distribution $P_{XY}$. Doing
so then naturally defines the density matrix $\rho_{XYB} = \sum_{x,y}
P_{XY}(x,y) \proj{x}\otimes\proj{y}\otimes \rho_B^x$, and thus also
the density matrix $\rho_{YB} = \tr_X(\rho_{XYB}) = \sum_{y}
P_Y(y)\proj{y}\otimes\big( \sum_x P_{X|Y}(x|y) \rho_B^x\big)$. If the
meaning is clear from the context, we tend to slightly abuse notation
and write the latter also as $\rho_{YB} = \sum_{y}
P_Y(y)\proj{y}\otimes\rho_B^y$, i.e., understand $\rho_B^y$ as $\sum_x
P_{X|Y}(x|y) \rho_B^x$. Throughout, we write $\id$ for the identity
matrix of appropriate dimension.

\subsection{Distance and Entropy Measures for Quantum States}\label{sec:entropies}

We recall some definitions from \cite{Renner05}. Let $\rho_{XB} \in
\dens{\cH_X \otimes \cH_B}$. Although the following definitions make
sense (and are defined in~\cite{Renner05}) for arbitrary $\rho_{XB}$,
we may assume $\rho_{XB}$ to be normalized\footnote{For a
  non-normalized $\rho_{XB}$, there is a normalizing
  $1/\tr(\rho_{XB})$-factor in the definition of collision-entropy.
  Also note that $\tr(\sigma^{-1/2} \rho \sigma^{-1/2}) = \tr(\rho
  \sigma^{-1})$ for any invertible $\sigma$. } and to be classical on
$\cH_X$.

\begin{definition}
The
{\em $L_1$-distance from uniform} of $\rho_{XB}$ given $B$ is defined by
\[ 
d(\rho_{XB} | B ) \assign \| \rho_{XB} - \rho_U \otimes \rho_B
\|_1 = \tr\bigl(|\rho_{XB} - \rho_U \otimes \rho_B|\bigr) 
\]
where $\rho_U \assign \frac{1}{\dim(\cH_X)} \id$ is the fully mixed 
state on $\cH_X$ and $|A| \assign \sqrt{A^\dag A}$ is the positive square
root of $A^\dag A$ (where $A^\dag$ is the complex-conjugate transpose of $A$).
\end{definition}
If $\rho_{XB}$ is classical on $\cH_X$, then $d(\rho_{XB}|B) = 0$ if and only if $X$ is uniformly distributed and $\rho^x_B$ does not depend on $x$, which in particular implies
that no information on $X$ can be learned by observing system $\cH_B$. Furthermore, if $d(\rho_{XB}|B) \leq \eps$ then the real system $\rho_{XB}$ ``behaves'' as the ideal
system $\rho_U \otimes \rho_B$ except with probability~$\varepsilon$ in that for any evolution of the system no
observer can distinguish the real from the ideal one with advantage
greater than~$\eps$~\cite{RK05}. 

\begin{definition}\label{def:entropies}
The {\em collision-entropy} and the {\em min-entropy} of
$\rho_{XB}$ {\em relative to} a normalized and invertible $\sigma_B \in \dens{\cH_B}$ are defined by
\begin{align*}
\H_2(\rho_{XB}|\sigma_B) &\assign -\log 
\trace{ \left( (\id \otimes \sigma_B^{-1/4}) \: \rho_{XB} \: (\id \otimes \sigma_B^{-1/4}) \right)^2 } \\
&= -\log 
  \sum_x P_X(x)^2 \trace{ \left(\sigma_B^{-1/4} \: \rho^x_B \: \sigma_B^{-1/4} \right)^2 } \qquad\text{and} \\[0.5ex]
\Hoo(\rho_{XB}|\sigma_B) &\assign -\log 
\lambda_\text{max}{ \left( (\id \otimes \sigma_B^{-1/2}) \:
    \rho_{XB} \: (\id \otimes \sigma_B^{-1/2}) \right) } \\
&= -\log 
\max_x \lambda_\text{max}{\left( P_X(x) \: \sigma_B^{-1/2} \: \rho_{B}^x \: \sigma_B^{-1/2} \right) } \, ,
\end{align*}
respectively, where $\lambda_\text{max}(\cdot)$ denotes the largest eigenvalue of the argument. 
The {\em collision-entropy} and the {\em min-entropy} of
$\rho_{XB}$ {\em given} $\cH_B$ are defined by
$$
\H_2(\rho_{XB}|B) \assign \sup_{\sigma_B} \H_2(\rho_{XB}|\sigma_B)
\qquad\text{and}\qquad
\Hoo(\rho_{XB}|B) \assign \sup_{\sigma_B} \Hoo(\rho_{XB}|\sigma_B)
$$
respectively, 
where the supremum ranges over all normalized $\sigma_B \in
\dens{\cH_B}$. 
\end{definition}
Note that without loss of generality, the supremum over $\sigma_B$ can
be restricted to the set of normalized \emph{and invertible} states
$\sigma_B$ which is dense in the set of normalized states in
$\dens{\cH_B}$.
Note furthermore that it is not clear, neither in the classical nor in
the quantum case, what the ``right'' way to define conditional
collision- or min-entropy is, and as a matter of fact, it depends on
the context which version serves best.  An alternative way to define
the collision- and min-entropy of $\rho_{XB}$ given $\cH_B$ would be
as $\tilde{\H}_2(\rho_{XB}|B) \assign \H_2(\rho_{XB}|\rho_B)$ and
$\tilde{\H}_\infty(\rho_{XB}|B) \assign \Hoo(\rho_{XB}|\rho_B)$. For a
density matrix $\rho_{XY}$ that is classical on $\cH_X$ and $\cH_Y$,
it is easy to see that $\tilde{\H}_2(\rho_{XY}|Y) = -\log \sum_y
P_Y(y) \sum_x P_{X|Y}(x|y)^2$, i.e., the negative logarithm of the average
conditional collision probability, and
\smash{$\tilde{\H}_\infty(\rho_{XY}|Y) = -\log \max_{x,y}
  P_{X|Y}(x|y)$}, i.e., the negative logarithm of the maximal conditional
guessing probability. These notions of classical conditional
collision- and min-entropy are commonly used in the literature,
explicitly (see~e.g.~\cite{RW05,DFSS06}) or implicitly (as~e.g.\
in~\cite{BBCM95}). We stick to Definition~\ref{def:entropies} because
it leads to stronger results, in that asking $\H_2(\rho_{XB}|B)$ to be
large is a weaker requirement than asking $\tilde{\H}_2(\rho_{XB}|B)$
to be large, as obviously $\H_2(\rho_{XB}|B) \geq
\tilde{\H}_2(\rho_{XB}|B)$, and similarly for the min-entropy.

\section{The New Randomness-Extraction Result}\label{sec:main}

We start by recalling the definition of a $\delta$-biased random variable and of a $\delta$-biased family of random variables~\cite{NN90,DS05}.

\begin{definition}
The {\em bias} of a random variable $A$, with respect to $\alpha \in \set{0,1}^n$, is defined as
$$
\bias(A) := \sum_a P_A(a) (-1)^{\alpha \cdot a} = 2\big( P[\alpha \!\cdot\! A \!=\! 1] - \textstyle\frac12 \big) \, ,
$$
and $A$ is called {\em $\delta$-biased} if $\bias(A)\leq\delta$ for
all non-zero $\alpha \in \set{0,1}^n$. A family of random
variables $\set{A_i}_{i \in \cI}$ over $\set{0,1}^n$ is called {\em
  $\delta$-biased} if, for all $\alpha \neq 0$,
$$
\sqrt{\E_{i \from \cI}\bigl[\bias(A_i)^2\bigr]} \leq \delta
$$
where the expectation is over a $i$ chosen uniformly at random from $\cI$.
\end{definition}
Note that by Jensen's inequality, $\E_{i \from \cI}[\bias(A_i)] \leq \delta$ for all
non-zero $\alpha$ is a necessary (but not sufficient) condition for
$\set{A_i}_{i \in \cI}$ to be $\delta$-biased. In case though the family
consists of only one member, then it is $\delta$-biased if and only if its only member is.

Our main theorem states that if $\set{A_i}_{i \in \cI}$ is $\delta$-biased for a small $\delta$, and if an adversary's conditional entropy $\H_2(\rho_{XB} | B)$ on a string $X \in \set{0,1}^n$ is large enough, then masking $X$ with $A_i$ for a random but known $i$ gives an essentially random string.

\begin{theorem}\label{thm:main}
Let the density matrix $\rho_{XB} \in \dens{\cH_X \otimes \cH_B}$ be classical on $\cH_X$ with $X \in \set{0,1}^n$. 
  Let $\set{A_i}_{i \in \cal I}$ be a $\delta$-biased family of random
  variables over $\set{0,1}^{n}$, and let $I$ be uniformly and independently distributed over $\cI$. Then
\[ d\big(\rho_{(A_I \oplus X) B I} \big| B I\big) \leq \delta \cdot 
2^{- \frac12(\H_2(\rho_{XB} | B) - n)}. 
\]
\end{theorem}
By the inequalities
$$
\Hoo(X) - \log\dim(\cH_B) \leq \Hoo(\rho_{XB}|B) \leq \H_2(\rho_{XB}|B) \, ,
$$
proven in~\cite{Renner05}, Theorem~\ref{thm:main} may also be expressed in terms of conditional min-entropy $\Hoo(\rho_{XB}|B)$ or in terms of classical min-entropy of $X$ minus the size of the quantum state (i.e.~the number of qubits). 
If $B$ is the ``empty'' quantum state, i.e., $\log\dim(\cH_B) = 0$, then Theorem~\ref{thm:main} coincides with Lemma~4 of~\cite{DS05}. 
Theorem~\ref{thm:main} also holds, with a corresponding normalization factor, for non-normalized operators, from which it follows that it can also be expressed in terms of the {\em smooth} conditional min-entropy $\Hoo[\eps](\rho_{XB}|B)$, as defined in~\cite{Renner05}, as $d(\rho_{(A_I \oplus X) B I} | B I) \leq 2 \eps + \delta \cdot 2^{-\frac12(\Hoo[\eps](\rho_{XB} | B) - n)}. $

\section{The Proof}\label{sec:proof}

We start by pointing out some elementary observations regarding the Fourier transform over the hypercube. In particular, we can extend the Convolution theorem and Parseval's identity to the case of matrix-valued functions. Further properties of the Fourier transform (with a different normalization) of matrix-valued functions over the hypercube have recently been established by Ben-Aron, Regev and de Wolf \cite{BRW07}. In Section~\ref{sec:Renato}, we introduce and recall a couple of properties of the $L_2$-distance from uniform. The actual proof of Theorem~\ref{thm:main} is given in Section~\ref{sec:actproof}. 

\subsection{Fourier Transform and Convolution}

For some fixed positive integer $d$, consider the complex vector space
$\cal MF$ of all functions $M:\set{0,1}^n \to \C^{d \times d}$. 
The {\em convolution} of two such matrix-valued functions $M,N \in
\cal MF$ is the matrix-valued function 
$$
M*N: x \mapsto \sum_y M(y)N(x-y)
$$
and the 
{\em Fourier transform} of a matrix-valued function $M \in \cal MF$ is the matrix-valued function 
$$
\F(M): \alpha \mapsto 2^{-n/2}\sum_x (-1)^{\alpha \cdot x} M(x)
$$
where $\alpha \cdot x$ denotes the standard inner product modulo~$2$.
Note that if $X$ is a random variable with distribution $P_X$ and $M$ is the matrix-valued function \mbox{$x \mapsto P_X(x) \cdot \id$}, then 
$$
\F(M)(\alpha) = 2^{-n/2} \cdot \bias(X) \cdot \id \, .
$$ 
The {\em Euclidean} or \emph{$L_2$}-norm of a matrix-valued function $M \in \cal MF$ is given by
$$
\Norm{M} := \sqrt{\tr\bigg(\sum_x M(x)^\dag M(x)\bigg)} 
$$
where $M(x)^\dag$ denotes the complex-conjugate transpose of the matrix $M(x)$.\footnote{We will only deal with Hermitian matrices $M(x)$ where $\Norm{M} = \sqrt{\tr\big(\sum_x M(x)^2 \big)}$. }

The following two properties known as Convolution Theorem and Parseval's Theorem are straightforward to prove (see Appendix~\ref{app:Fourier}). 
\begin{lemma}\label{lemma:Fourier}
For all $M,N \in \cal MF$: 
$$
\F(M*N) = 2^{n/2} \cdot\F(M) \cdot \F(N)
\qquad\text{and}\qquad
\Norm{\F(M)} = \Norm{M} \, .
$$
\end{lemma}

\subsection{The $L_2$-Distance from Uniform}\label{sec:Renato}

The following lemmas together with their proofs can be found in~\cite{Renner05}. Again, we restrict ourselves to the case where $\rho_{XB}$ and $\sigma_B$ are normalized and $\rho_{XB}$ is classical on $X$, whereas the claims hold (partly) more generally. 

\begin{definition}
Let $\rho_{XB} \in \dens{\cH_X \otimes \cH_B}$ and $\sigma_B \in
\dens{\cH_B}$. Then the conditional $L_2$-distance from uniform of
$\rho_{XB}$ relative to $\sigma_B$ is 
\begin{align*}
d_2(\rho_{XB} | \sigma_B) &\assign 
\trace{ \left( (\id \otimes \sigma_B^{-1/4})(\rho_{XB}-\rho_U \otimes
  \rho_B)(\id \otimes \sigma_B^{-1/4}) \right)^2  },
\end{align*}
where $\rho_U \assign \frac{1}{\dim(\cH_X)} \id$ is the fully mixed
state on $\cH_X$.
\end{definition}

\begin{lemma} \label{lem:l1withl2}
Let $\rho_{XB} \in \dens{\cH_X \otimes \cH_B}$. Then, for any normalized
$\sigma_B \in \dens{\cH_B}$,
\[
d(\rho_{XB} | B) \leq \sqrt{\dim(\cH_X)} \sqrt{d_2(\rho_{XB}|\sigma_B)}.
\]
\end{lemma}

\begin{lemma}\label{lem:d2explicit}
  Let $\rho_{XB} \in \dens{\cH_X \otimes \cH_B}$ be classical on
  $\cH_X$ with $X \in \cal X$, and let $\rho^x_{B}$ be the corresponding normalized
  conditional operators. Then, for any $\sigma_B \in \dens{\cH_B}$
\[ d_2(\rho_{XB} | \sigma_B) = \sum_x \trace{(\sigma_B^{-1/4} P_X(x) \rho^x_B
  \sigma_B^{-1/4})^2} - \frac{1}{|\cX|} \trace{(\sigma_B^{-1/4} \rho_B \sigma_B^{-1/4})^2}.
\]
\end{lemma}

\subsection{Proof~Theorem~\ref{thm:main}}\label{sec:actproof}

Write $D_i = A_i \oplus X$ and $D_I = A_I \oplus X$. 
Since $\rho_{D_I B I} = \frac{1}{|{\cal I}|}\sum_i \rho_{D_I B}^i \otimes \proj{i} = \frac{1}{|{\cal I}|}\sum_i \rho_{D_i B} \otimes \proj{i}$, and similar for $\rho_{B I}$, it follows that 
the $L_1$-distance from uniform can be written as an expectation over the random choice of $i$ from $\cI$. Indeed
\begin{align*}
d(&\rho_{D_I B I} | B I) = \frac{1}{|{\cal I}|} \tr\Bigl(\Big|\sum_i (\rho_{D_i B}  - \rho_U \otimes \rho_{B}) \otimes \proj{i}\Big|\Bigr) \\ 
&= \frac{1}{|{\cal I}|}\sum_i \tr\bigl(\big| \rho_{D_i B}  - \rho_U \otimes \rho_{B} \big|\bigr) 
= \frac{1}{|{\cal I}|}\sum_i d(\rho_{D_i B} | B) = \E_{i \from \cI}\big[ d(\rho_{D_i B} | B) \big] \, .
\end{align*}
where the second equality follows from the block-diagonal form of the matrix. 
With Lemma~\ref{lem:l1withl2}, the term in the expectation can be bounded in terms of the $L_2$-distance from uniform, that is, for any normalized $\sigma_B \in \dens{\cH_B}$, 
\begin{align*}
d(\rho_{D_I B I} |B I) &\leq  \sqrt{ 2^n } \: \E_{i \from \cI} \Bigl[\sqrt{d_2(\rho_{D_i B} | \sigma_B)} \Bigr] \leq 2^{n/2} \sqrt{ \E_{i \from \cI} \bigl[ d_2(\rho_{D_i B} | \sigma_B) \bigr] }
\end{align*}
where the second inequality is Jensen's inequality. 
By Lemma~\ref{lem:d2explicit}, we have for the $L_2$-distance
\begin{align} \label{eq:firsteq} \begin{split}
&d_2(\rho_{D_i B} | \sigma_B)\\
&= \trace{ \sum_d  (\sigma_B^{-1/4} \: P_{D_i}(d) \rho^d_B \: \sigma_B^{-1/4})^2 } - \frac{1}{2^n} \trace{ (\sigma_B^{-1/4}\, \rho_B \, \sigma_B^{-1/4})^2} \, .
\end{split} \end{align}
Note that 
\begin{align*}
P_{D_i}(d) &\rho^d_B = P_{D_i}(d) \sum_x P_{X|D_i}(x | d) \rho^x_B = \sum_x P_{X D_i}(x,d) \rho^x_B \\
&= \sum_x P_{X A_i}(x,d \oplus x) \rho^x_B = \sum_x P_X(x) P_{A_i}(d \oplus x) \rho^x_B
\end{align*}
so that the first term on the right-hand side of~\eqref{eq:firsteq} can be written as
\begin{align*} 
&\trace{ \sum_d (\sigma_B^{-1/4} \: P_{D_i}(d) \rho^d_B \:
  \sigma_B^{-1/4})^2 }\\
&=\trace{ \sum_d \bigg(\sum_x P_X(x) \sigma_B^{-1/4} \: \rho^x_B \: \sigma_B^{-1/4}   P_{A_i}(d \oplus x) \bigg)^2 } \, .
\end{align*}
The crucial observation now is that the term that is squared on the
right side is the convolution of the two matrix-valued functions $M: x
\mapsto P_X(x) \sigma_B^{-1/4} \: \rho^x_B \: \sigma_B^{-1/4}$ and $N:
x \mapsto P_{A_i}(x) \id$, and the whole expression equals
$\Norm{M*N}^2$. Thus, using Lemma~\ref{lemma:Fourier} 
we get
\begin{align} \begin{split} \label{eq:alphanotzero}
  \tr\Biggl( \sum_d (&\sigma_B^{-1/4} \: P_{D_i}(d) \rho^d_B \:
  \sigma_B^{-1/4})^2 \Biggr) = \Norm{M*N}^2 = \Norm{\F(M*N)}^2\\
  &= \Norm{2^{n/2} \cdot \F(M)\cdot\F(N)}^2 = 2^{n} \trace{ \sum_\alpha \big(\F(M)(\alpha) \F(N)(\alpha)\big)^2 } \\
  &=
  \frac{1}{2^{n}} \trace{ (\sigma_B^{-1/4} \: \rho_B \: \sigma_B^{-1/4})^2 } +
  \trace{ \sum_{\alpha\neq 0} \, \F(M)(\alpha)^2 \, \bias(A_i)^2 } \, ,
\end{split} \end{align}
where the last equality uses
$$
\F(M)(0) = 2^{-n/2} \sum_x P_X(x) \sigma_B^{-1/4} \: \rho^x_B \:
\sigma_B^{-1/4} = 2^{-n/2} \sigma_B^{-1/4} \: \rho_B \:
\sigma_B^{-1/4}
$$ 
as well as
$$
\F(N)(0) =  2^{-n/2} \sum_x P_{A_i}(x) \id = 2^{-n/2} \id 
\quad\text{and}\quad
\F(N)(\alpha) = 2^{-n/2} \cdot \bias(A_i)\id \, .
$$
Substituting \eqref{eq:alphanotzero} into \eqref{eq:firsteq} gives
$$
d_2(\rho_{D_i B} | \sigma_B) = \trace{ \sum_{\alpha\neq 0} \F(M)(\alpha)^2 \, \bias(A_i)^2 } \, .
$$
Using the linearity of the expectation and trace, and using the bound
on the expected square-bias, we get
\begin{align*}
\E_{i \from \cI} \big[ &d_2(\rho_{D_i B} | \sigma_B) \big] 
\leq \delta^2 \tr\biggl( \sum_{\alpha\neq 0} \F(M)(\alpha)^2 \biggr) \leq \delta^2\tr\biggl( \sum_{\alpha} \F(M)(\alpha)^2 \biggr) \\
&= \delta^2 \Norm{\F(M)}^2  = \delta^2 \Norm{M}^2 
= \delta^2 \sum_x \trace{ P_X(x)^2 (\sigma_B^{-1/4} \:  \rho^x_B \: \sigma_B^{-1/4})^2 } \\
&= \delta^2 2^{-\H_2(\rho_{XB} | \sigma_B)} \, ,
\end{align*}
where the second inequality follows because of
$$\tr\bigl(\F(M)(0)^2\bigr) = 2^{-n} \tr\bigl( (\sigma_B^{-1/4} \:
\rho_B \: \sigma_B^{-1/4} )^2 \bigr) \geq 0 \, .$$  Therefore,
$$
d(\rho_{D_I B I} | B I) \leq 2^{n/2} \sqrt{ \E_{i \from \cI} \bigl[ d_2(\rho_{D_i B} | \sigma_B) \bigr] } \leq \delta \cdot 2^{-\frac12(\H_2(\rho_{XB} | \sigma_B)-n)}
$$
and the assertion follows from the definition of $H_2(\rho_{XB}|B)$
because $\sigma_B$ was arbitrary. 
\qed

\section{Application I: Entropic Security}
\label{sec:AppI}

Entropic security is a relaxed but still meaningful security
definition for (information-theoretically secure) encryption that
allows to circumvent Shannon's pessimistic result, which states that
any perfectly secure encryption scheme requires a key at least as long
as the message to be encrypted. Entropic security was introduced by
Russell and Wang~\cite{RW02}, and later more intensively investigated
by Dodis and Smith~\cite{DS05tcc}. Based on our result, and in
combination with techniques from~\cite{DS05tcc}, we show how to
achieve entropic security against quantum adversaries. We would like
to stress that in contrast to perfect security e.g.~when using the
one-time-pad, entropic security does {\em not} a priori protect
against a quantum adversary.

Informally, entropic security is defined as follows.
An encryption scheme is entropically secure if no adversary can obtain any information on the message $M$ from its ciphertext $C$ (in addition to what she can learn from scratch), provided the message $M$ has enough uncertainty from the adversary's point of view. 
The impossibility of obtaining any information on $M$ is formalized by requiring that any adversary that can compute $f(M)$ for some function $f$ when given $C$, can also compute $f(M)$ {\em without}~$C$ (with similar success probability). 
A different formulation, which is named {\em indistinguishability}, is to require that there exists a random variable $C'$, independent of $M$, such that $C$ and $C'$ are essentially identically distributed. It is shown in~\cite{DS05tcc}, and in~\cite{Des07} for the case of a {\em quantum} message, that the two notions are equivalent if the adversary's information on $M$ is classical.
In recent work, Desrosiers and Dupuis proved this equivalence to hold also for an adversary with quantum information~\cite{DD07}. 

The adversary's uncertainty on $M$ is formalized, for a {\em classical} (i.e.~non-quantum) adversary, by the {\em min-entropy} $\Hoo(M|V\!=\!v)$ (or, alternatively, the collision-entropy) of $M$, conditioned on the value $v$ the adversary's view $V$ takes on.
We formalize this uncertainty for a quantum adversary in terms of the quantum version of conditional min- or actually collision-entropy, as introduced in Section~\ref{sec:entropies}. 

\begin{definition}\label{def:q-ind}
We call a (possibly randomized) encryption scheme $E: \K \times
{\cal M} \to {\cal C}$ {\em $(t,\eps)$-quantum-indistinguishable} if
there exists a random variable $C'$ over $\cal C$ such that for any
normalized $\rho_{MB} \in \dens{\cH_M \otimes \cH_B}$ which is classical on $\cH_M$ with $M
\in \cal M$ and $\H_2(\rho_{MB}|B) \geq t$, we have that
$$
\big\|\rho_{E(K,M) B} - \rho_{C'} \otimes \rho_B \big\|_1 \leq \eps \, ,
$$
where $K$ is uniformly and independently distributed over $\K$. 
\end{definition} 
Note that in case of an ``empty'' state $B$, our definition coincides
with the indistinguishability definition from~\cite{DS05tcc} (except
that we express it in collision- rather than min-entropy). 

Theorem~\ref{thm:main}, with $\cI = \set{i_\nord}$ and $A_{i_\nord} = K$, immediately gives a generic construction for a quantum-indistinguishable encryption scheme (with $C'$ being uniformly distributed). Independently, this result was also obtained in~\cite{DD07}.
\begin{theorem}
Let $\K \subseteq \set{0,1}^n$ be such that the uniform distribution $K$ over $\K$ is $\delta$-biased. Then the encryption scheme $E: \K \times \set{0,1}^n \to \set{0,1}^n$ with $E(k,m) = k \oplus m$ is $(t,\eps)$-quantum-indistinguishable with \smash{$\eps = \delta \cdot 2^{\frac{n - t}{2}}$}. 
\end{theorem}
Alon \etal~\cite{AGHP90} showed how to construct subsets $\K \subseteq \set{0,1}^n$ of
size $|\K| = O(n^2/\delta^2)$ such that the uniform distribution $K$
over $\K$ is $\delta$-biased and elements in $\K$ can be efficiently sampled. With the help of this construction, we
get the following result, which generalizes the bound on the key-size obtained
in~\cite{DS05tcc} to the quantum setting.
\begin{corollary}\label{cor:Ind}
For any $\eps \geq 0$ and $0 \leq t \leq n$, there exists a
$(t,\eps)$-quantum-indistinguishable encryption scheme encrypting
$n$-bit messages with key length $\ell = \log|\K| = n - t + 2\log(n) + 2\log(\frac{1}{\eps}) + O(1)$. 
\end{corollary}
In the language of extractors, defining a $(t,\eps)$-{\em quantum extractor} in the natural way as follows, Corollary~\ref{cor:Ind} translates to Corollary~\ref{cor:WeakExtr} below. 
\begin{definition}\label{def:Extr} 
A function $E: {\cal J} \times {\cal X} \to\set{0,1}^m$ is called a {\em
  $(t,\eps)$-weak quantum extractor} if
$
d(\rho_{E(J,X) B}|B) \leq \eps,
$
and a \emph{$(t,\eps)$-strong quantum extractor} if
$
d(\rho_{E(J,X) J B}|JB) \leq \eps
$
for any
normalized $\rho_{XB} \in \dens{\cH_X \otimes \cH_B}$ which is classical on $\cH_X$ with $X
\in \cal X$ and $\H_2(\rho_{XB}|B) \geq t$,
and where $J$ is uniformly and independently distributed over~$\cal J$. 
\end{definition} 
\begin{corollary}\label{cor:WeakExtr}
For any $\eps\geq 0$ and $0 \leq t \leq n$, there exists a $(t,\eps)$-weak quantum extractor with $n$-bit output and seed length $\ell = \log|\K| = n - t + 2\log(n) + 2\log(\frac{1}{\eps}) + O(1)$. 
\end{corollary}

\section{Application II: Private Error Correction}\label{sec:AppII}

Consider the following scenario.  Two parties, Alice and Bob, share a
common secret key $K$. Furthermore, we assume a ``random source" which
can be queried by Alice and Bob so that on identical queries it
produces identical outputs. In particular, when Alice and Bob both
query the source on input $K$, they both obtain the same ``raw key''
$X \in \set{0,1}^n$. We also give an adversary Eve access to the
source. She can obtain some (partial) information on the source and
store it possibly in a quantum state $\rho_Z$. However, we assume she
has some uncertainty about~$X$, because due to her ignorance of $K$,
she is unable to extract ``the right" information from the source.
Such an assumption of course needs to be justified in a specific
implementation. Specifically, we require that $\Hoo(\rho_{XKZ}|KZ)$ is
lower bounded, i.e., Eve has uncertainty in $X$ even if at some later
point she learns $K$ but only the source has disappeared in the meantime.

Such a scenario for instance arises in the bounded-storage
model~\cite{Maurer90,ADR02} (though with classical Eve), when $K$ is
used to determine which bits of the long randomizer Alice and Bob
should read to obtain $X$, or in a quantum setting when Alice sends
$n$ qubits to Bob and $K$ influences the basis in which Alice prepares
them respectively Bob measures them.

In this setting, it is well-known how to transform by public
(authenticated) communication the weakly-secure raw key $X$ into a
fully secure key $S$: Alice and Bob do privacy amplification, as shown
in~\cite{HILL99,BBCM95} in case of a classical Eve, respectively as
in~\cite{RK05,Renner05} in case of a quantum Eve. Indeed, under the
above assumptions on the entropy of $X$, privacy amplification
guarantees that the resulting key $S$ looks essentially random for Eve
even given $K$. This guarantee implies that $S$ can be used, say, as a
one-time-pad encryption key, but it also implies that if Eve
learns $S$, she still has essentially no information on $K$, and thus
$K$ can be safely re-used for the generation of a new key $S$.

Consider now a more realistic scenario, where due to noise or
imperfect measurements Alice's string $X$ and Bob's string $X'$ are
close but not exactly equal. There are standard techniques to do
error correction (without giving Eve too much information on $X$):
Alice and Bob agree on a suitable error-correcting code $\cal C$,
Alice samples a random codeword $C$ from $\cal C$ and sends $Y = X \oplus C$
to Bob, who can recover $X$ by decoding $C' = Y \oplus X'$ to the nearest
codeword $C$ and compute $X = Y \oplus C$. Or equivalently, in case of a
linear code, Alice can send the syndrome of $X$ to Bob, which
allows Bob to recover $X$ in a similar manner. If Eve's entropy in $X$ is significantly
larger than the size of the syndrome, then one can argue that privacy
amplification still works and the resulting key $S$ is still (close
to) random given Eve's information (including the syndrome) and
$K$. Thus, $S$ is still a secure key. However, since $X$ depends on
$K$, and the syndrome of $X$ depends on $X$, the syndrome of $X$ may
give information on $K$ to Eve, which makes it insecure to re-use $K$.
A common approach to deal with this problem is to use part of $S$ as
the key $K$ in the next session. Such an approach not only creates a lot of
inconvenience for Alice and Bob in that they now have to be stateful
and synchronized, but in many cases Eve can prevent Alice and Bob
from agreeing on a secure key $S$ (for instance by blocking the last
message) while nevertheless learning information on $K$, and thus Eve
can still cause Alice and Bob to run out of key material.

In~\cite{DS05}, Dodis and Smith addressed this problem and proposed an
elegant solution in case of a classical Eve. They constructed a family
of codes which not only allow to efficiently correct errors, but at
the same time also serve as randomness extractors. More precisely,
they show that for every $0 < \lambda <1$, there exists a family
$\set{{\cal C}_j}_{j \in \cal J}$ of binary linear codes of length
$n$, which allows to efficiently correct a constant fraction of
errors, and which is $\delta$-biased for $\delta < 2^{-\lambda n/2}$.
The latter is to be understood that the family $\set{C_j}_{j \in \cal
  J}$ of random variables, where $C_j$ is uniformly distributed over
${\cal C}_j$, is $\delta$-biased for $\delta < 2^{-\lambda n/2}$.
Applying Lemma~4 of~\cite{DS05} (the classical version of
Theorem~\ref{thm:main}) implies that $C_j \oplus X$ is close to random
for any $X$ with large enough entropy, given $j$. Similarly, applying
our Theorem~\ref{thm:main} implies the following.

\begin{theorem}\label{thm:ErrorCorrection}
For every $0 < \lambda <1$ there exists a family $\set{{\cal C}_j}_{j
  \in \cal J}$ of binary linear codes of length $n$ which allows to
efficiently correct a constant fraction of errors, and such that for
any density matrix $\rho_{XB} \in \dens{\cH_X \otimes \cH_B}$ which is
classical on $\cH_X$ with $X \in \set{0,1}^n$ and $\H_2(\rho_{XB}|B)
\geq t$, it holds that
$$
d\big(\rho_{(C_J \oplus X) B J}\big|B J \big) \leq 2^{-\frac{t - (1-\lambda)n}{2}} \, ,
$$
where $J$ is uniformly distributed over $\cal J$ and $C_J$ is uniformly distributed over ${\cal C}_J$. 
\end{theorem}
Using a random code from such a family of codes allows to do error
correction in the noisy setting described above without leaking
information on $K$ to Eve: By the chain rule
\cite[Sect.~3.1.3]{Renner05}, the assumed lower bound on
$\Hoo(\rho_{XKZ}|KZ)$ implies a lower bound on
$\Hoo(\rho_{XSKZG}|SKZG)$ (essentially the original bound minus the
bit length of $S$), where $G$ is the randomly chosen universal hash
function used to extract $S$ from $X$. Combining systems $S,K,Z$
and $G$ into system $B$, Theorem~\ref{thm:ErrorCorrection} 
implies that $\rho_{(C_J\oplus X)SKZGJ} \approx
\frac{1}{2^n}\I\otimes\rho_{SKZGJ}$.  From standard privacy
amplification follows that $\rho_{SKZGJ} \approx
\frac{1}{2^\ell}\I\otimes\rho_{KZGJ}$. Using the independence of $K, G, J$
(from $Z$ and from each other), we obtain $\rho_{(C_J\oplus X)SKZGJ} \approx
\frac{1}{2^n}\I\otimes\frac{1}{2^\ell}\I\otimes\rho_K \otimes \rho_Z
\otimes \rho_G \otimes \rho_J$.  This in particular implies that $S$
is a secure key (even when $K$ is given to Eve) and that $K$
is still ``fresh" and can be safely re-used (even when $S$ is additionally
given to Eve).

Specifically, our private-error-correction techniques allow to add
robustness against noise to the bounded-storage model in the
presence of a quantum attacker as considered in~\cite{KR07}, without
the need for updating the common secret key. The results
of~\cite{KR07} guarantee that the min-entropy of the sampled substring is
lower bounded given the attacker's quantum information and hence,
security follows as outlined above.
Furthermore,
in~\cite{DFSS07} the above
private-error-correction technique is an essential ingredient to add robustness against
noise but also to protect against man-in-the-middle attacks in new
quantum-identification and quantum-key-distribution schemes in the
bounded-quantum-storage model.

In the language of extractors, we get the following result for arbitrary, not necessarily efficiently decodable, binary linear codes. 
\begin{corollary}
Let $\set{{\cal C}_j}_{j \in \cal J}$ be a $\delta$-biased family of binary linear $[n,k,d]_2$-codes. For any $j \in {\cal J}$, let $G_j$ be a generator matrix for the code ${\cal C}_j$ and let $H_j$ be a corresponding parity-check matrix. Then $E: {\cal J} \times \set{0,1}^n \to \set{0,1}^{n-k}$, $(j,x) \mapsto H_j x$ is a $(t,\eps)$-strong quantum extractor with $\eps = \delta \cdot 2^{\frac12(n-t)}$. 
\end{corollary}
This result gives rise to new privacy-amplification techniques, beyond
using universal$_2$ hashing as in~\cite{RK05} or one-bit extractors as
in~\cite{KT06}. Note that using arguments from~\cite{DS05}, it is easy
to see that the condition that $\set{{\cal C}_j}_{j \in \cal J}$ is
$\delta$-biased and thus the syndrome function $H_j$ is a good strong
extractor, is equivalent to requiring that $\set{G_j}_{j \in \cal J}$
seen as family of (encoding) functions is $\delta^2$-almost
universal$_2$~\cite{WC81,Stinson92}.

For a family of binary linear codes $\set{{\cal C}_j}_{j \in \cal J}$, another equivalent condition for $\delta$-bias of $\set{{\cal C}_j}_{j \in \cal J}$ is to require that for all non-zero $\alpha$, $\Pr_{j \in {\cal J}}[ \alpha \in {\cal C}_j^{\perp} ] \leq \delta^2$, i.e.~that the probability that $\alpha$ is in the dual code of ${\cal C}_j$ is upper bounded by $\delta^2$ \cite{DS05}. It follows that the family size $|\cal J|$ has to be exponential in $n$ to achieve an exponentially small bias $\delta$ and therefore, the seed length $\log|\cal J|$ of the strong extractor will be linear in $n$ as for the case of two-universal hashing.

\section{Conclusion}\label{sec:Concl}

We proposed a new technique for randomness extraction in the presence
of a quantum attacker. This is interesting in its own right, as up to
date only very few extractors are known to be secure against quantum
adversaries, much in contrast to the classical non-quantum case.  The
new randomness-extraction technique has various cryptographic
applications like entropically secure encryption, in the classical
bounded-storage model and the bounded-quantum-storage model, and in 
quantum key distribution. Furthermore, because of the wide range of
applications of classical extractors not only in cryptography but also
in other areas of theoretical computer science, we feel that our new
randomness-extraction technique will prove to be useful in other
contexts as well.

\section*{Acknowledgments}
We would like to thank Ivan Damg{\aa}rd, Renato Renner, and Louis
Salvail for helpful discussions and the anonymous referees for useful
comments.


\bibliographystyle{abbrv} 
\bibliography{qip,crypto,procs}

\begin{thebibliography}{10}

\bibitem{AGHP90}
N.~Alon, O.~Goldreich, J.~H{\aa}stad, and R.~Peralta.
\newblock Simple constructions of almost k-wise independent random variables.
\newblock In {\em 31st Annual IEEE Symposium on Foundations of Computer Science
  (FOCS)}, volume~II, pages 544--553, 1990.

\bibitem{AS04}
A.~Ambainis and A.~Smith.
\newblock Small pseudo-random families of matrices: Derandomizing approximate
  quantum encryption.
\newblock In K.~Jansen, S.~Khanna, J.~D.~P. Rolim, and D.~Ron, editors, {\em
  Approximation Algorithms for Combinatorial Optimization Problems, {APPROX}
  2004, and 8th International Workshop on Randomization and Computation,
  {RANDOM} 2004}, volume 3122 of {\em Lecture Notes in Computer Science}, pages
  249--260. Springer, 2004.

\bibitem{ADR02}
Y.~Aumann, Y.~Z. Ding, and M.~O. Rabin.
\newblock Everlasting security in the bounded storage model.
\newblock {\em {IEEE} Transactions on Information Theory}, 48(6):1668--1680,
  June 2002.

\bibitem{BRW07}
A.~{Ben-Aroya}, O.~Regev, and R.~de~Wolf.
\newblock A hypercontractive inequality for matrix-valued functions with
  applications to quantum computing.
\newblock {\tt http://arxiv.org/abs/0705.3806}, 2007.

\bibitem{BBCM95}
C.~H. Bennett, G.~Brassard, C.~Cr{\'e}peau, and U.~M. Maurer.
\newblock Generalized privacy amplification.
\newblock {\em {IEEE} Transactions on Information Theory}, 41:1915--1923, Nov.
  1995.

\bibitem{DFSS06}
I.~B. Damg{\aa}rd, S.~Fehr, L.~Salvail, and C.~Schaffner.
\newblock Oblivious transfer and linear functions.
\newblock In {\em Advances in Cryptology---CRYPTO~'06}, volume 4117 of {\em
  Lecture Notes in Computer Science}, pages 427--444. Springer, 2006.

\bibitem{DFSS07}
I.~B. Damg{\aa}rd, S.~Fehr, L.~Salvail, and C.~Schaffner.
\newblock Secure identification and {QKD} in the bounded-quantum-storage model.
\newblock In {\em Advances in Cryptology---CRYPTO~'07}, volume 4622 of {\em
  Lecture Notes in Computer Science}, pages 342--359. Springer, 2007.

\bibitem{Des07}
S.~P. Desrosiers.
\newblock Entropic security in quantum cryptography.
\newblock {\tt http://arxiv.org/abs/quant-ph/0703046}, 2007.

\bibitem{DD07}
S.~P. Desrosiers and F.~Dupuis.
\newblock Quantum entropic security and approximate quantum encryption.
\newblock {\tt http://arxiv.org/abs/0707.0691}, July 5, 2007.

\bibitem{DN06}
P.~A. Dickinson and A.~Nayak.
\newblock Approximate randomization of quantum states with fewer bits of key.
\newblock In {\em Quantum Computing: Back Action 2006}, volume 864 of {\em
  American Institute of Physics Conference Series}, pages 18--36, November
  2006.
\newblock {\tt quant-ph/0611033}.

\bibitem{DS05}
Y.~Dodis and A.~Smith.
\newblock Correcting errors without leaking partial information.
\newblock In {\em 37th Annual ACM Symposium on Theory of Computing (STOC)},
  pages 654--663, 2005.

\bibitem{DS05tcc}
Y.~Dodis and A.~Smith.
\newblock Entropic security and the encryption of high entropy messages.
\newblock In {\em Theory of Cryptography Conference (TCC)}, volume 3378 of {\em
  Lecture Notes in Computer Science}, pages 556--577. Springer, 2005.

\bibitem{GKKRW07}
D.~Gavinsky, I.~Kerenidis, J.~Kempe, R.~Raz, and R.~de~Wolf.
\newblock Exponential separations for one-way quantum communication complexity,
  with applications to cryptography.
\newblock In {\em 39th Annual ACM Symposium on Theory of Computing (STOC)},
  pages 516--525, 2007.
\newblock {\tt http://arxiv.org/abs/quant-ph/0611209}.

\bibitem{HILL99}
J.~H{\aa}stad, R.~Impagliazzo, L.~A. Levin, and M.~Luby.
\newblock A pseudorandom generator from any one-way function.
\newblock {\em SIAM Journal on Computing}, 28(4), 1999.

\bibitem{ILL89}
R.~Impagliazzo, L.~A. Levin, and M.~Luby.
\newblock Pseudo-random generation from one-way functions.
\newblock In {\em 21st Annual ACM Symposium on Theory of Computing (STOC)},
  pages 12--24, 1989.

\bibitem{KN05}
I.~Kerenidis and D.~Nagaj.
\newblock On the optimality of quantum encryption schemes.
\newblock {\em Journal of Mathematical Physics}, 47:092102, 2006.
\newblock {\tt http://arxiv.org/abs/quant-ph/0509169}.

\bibitem{KR07}
R.~K\"onig and R.~Renner.
\newblock Sampling of min-entropy relative to quantum knowledge.
\newblock In {\em Workshop on Quantum Information Processing (QIP 2008)}, 2007.

\bibitem{KT06}
R.~K\"onig and B.~M. Terhal.
\newblock The bounded storage model in the presence of a quantum adversary.
\newblock {\tt http://arxiv.org/abs/quant-ph/0608101}, 2006.

\bibitem{Maurer90}
U.~M. Maurer.
\newblock A provably-secure strongly-randomized cipher.
\newblock In {\em Advances in Cryptology---EUROCRYPT~'90}, volume 473 of {\em
  Lecture Notes in Computer Science}, pages 361--373. Springer, 1990.

\bibitem{NN90}
J.~Naor and M.~Naor.
\newblock Small-bias probability spaces: efficient constructions and
  applications.
\newblock In {\em 22nd Annual ACM Symposium on Theory of Computing (STOC)},
  pages 213--223, 1990.

\bibitem{NZ93}
N.~Nisan and D.~Zuckerman.
\newblock More deterministic simulation in logspace.
\newblock In {\em 25th Annual ACM Symposium on the Theory of Computing (STOC)},
  pages 235--244, 1993.

\bibitem{Renner05}
R.~Renner.
\newblock {\em Security of Quantum Key Distribution}.
\newblock PhD thesis, ETH Z\"urich (Switzerland), September 2005.
\newblock {\tt http://arxiv.org/abs/quant-ph/0512258}.

\bibitem{RK05}
R.~Renner and R.~K\"onig.
\newblock Universally composable privacy amplification against quantum
  adversaries.
\newblock In {\em Theory of Cryptography Conference (TCC)}, volume 3378 of {\em
  Lecture Notes in Computer Science}, pages 407--425. Springer, 2005.

\bibitem{RW05}
R.~Renner and S.~Wolf.
\newblock Simple and tight bounds for information reconciliation and privacy
  amplification.
\newblock In {\em Advances in Cryptology---ASIACRYPT~2005}, Lecture Notes in
  Computer Science, pages 199--216. Springer, 2005.

\bibitem{RW02}
A.~Russell and H.~Wang.
\newblock How to fool an unbounded adversary with a short key.
\newblock In {\em Advances in Cryptology---EUROCRYPT~'02}, volume 2332 of {\em
  Lecture Notes in Computer Science}, pages 133--148. Springer, 2002.

\bibitem{Shaltiel02}
R.~Shaltiel.
\newblock Recent developments in explicit constructions of extractors.
\newblock {\em Bulletin of the EATCS}, 77:67--95, 2002.

\bibitem{Smith07privcom}
A.~Smith, 2007.
\newblock Private communication.

\bibitem{Stinson92}
D.~R. Stinson.
\newblock Universal hashing and authentication codes.
\newblock In {\em Advances in Cryptology---CRYPTO~'91}, volume 576 of {\em
  Lecture Notes in Computer Science}, pages 74--85. Springer, 1991.

\bibitem{TaS96}
A.~Ta-Shma.
\newblock On extracting randomness from weak random sources.
\newblock In {\em 28th Annual ACM Symposium on the Theory of Computing (STOC)},
  pages 276--285, 1996.

\bibitem{WC81}
M.~N. Wegman and L.~Carter.
\newblock New hash functions and their use in authentication and set equality.
\newblock {\em J. Comput. Syst. Sci.}, 22(3):265--279, 1981.

\end{thebibliography}


\begin{appendix}

\section{Proof of Lemma~\ref{lemma:Fourier}}\label{app:Fourier}

Concerning the first claim,
\begin{align*}
\F(M*N)(\alpha) &= \frac{1}{2^{n/2}} \sum_x (-1)^{\alpha \cdot x} \sum_y M(y)N(x \oplus y)  \\
&= 2^{-n/2} \sum_y (-1)^{\alpha \cdot y}M(y) \sum_x (-1)^{\alpha\cdot(x \oplus y)} N(x \oplus y) \\
&= 2^{-n/2} \sum_y (-1)^{\alpha \cdot y}M(y) \sum_z (-1)^{\alpha\cdot z}N(z) \\
&= 2^{n/2} \cdot \F(M)(\alpha) \cdot \F(N)(\alpha) \, .
\end{align*}
The second claim is argued as follows.
\begin{align*}
\Norm{\F(M)}^2 &= \tr\bigg(\sum_\alpha \F(M)(\alpha)^\dag \F(M)(\alpha) \bigg)\\
&= 2^{-n} \tr\bigg(\sum_\alpha \Big( \sum_x (-1)^{\alpha \cdot x} M(x) \Big)^{\!*}\Big( \sum_{x'} (-1)^{\alpha \cdot x'} M(x') \Big) \bigg) \\
&= 2^{-n} \tr\bigg(\sum_{x,x'} M(x)^\dag M(x') \sum_{\alpha} (-1)^{\alpha \cdot (x \oplus x')} \bigg)\\
&= \tr\bigg(\sum_x M(x)^\dag M(x) \bigg) = \Norm{M}^2
\end{align*}
where the last equality follows from the fact that $\sum_{\alpha} (-1)^{\alpha \cdot y} = 2^n$ if $y = (0,\ldots,0)$ and $0$ otherwise. 
\qed

\end{appendix}

\end{document}